\documentclass{desyproc}

\begin{document}

\newcommand{\I}{{\rm i}}
\newcommand{\sD}{{\sf D}}
\newcommand{\sF}{{\sf F}}
\newcommand{\sL}{{\sf L}}
\newcommand{\sH}{{\sf H}}
\newcommand{\sM}{{\sf M}}
\newcommand{\sN}{{\sf N}}
\newcommand{\sW}{{\sf\Omega}}

\def\beq{\begin{equation}}
\def\eeq{\end{equation}}
\def\w{\omega}

\newcommand{\D}{{\rm d}}
\newcommand{\E}{{\rm e}}
\title{Flavor Stability Analysis of Supernova Neutrino Fluxes Compared with
Simulations}

\author{{\slshape Srdjan~Sarikas$^{1,2,3}$, Georg~G.~Raffelt$^3$}\\[1ex]
$^1$Dipartimento di Scienze Fisiche, Universit\` a di Napoli ``Federico II", 80126 Napoli, Italy \\
$^2$Istituto Nazionale di Fisica Nucleare --- Sezione di Napoli, 80126 Napoli, Italy \\
$^3$Max-Planck-Institut f\" ur Physik, F\" ohringer Ring 6, 80805
M\"unchen, Germany}


\acronym{HANSE 2011} 

\maketitle

\begin{abstract}
We apply a linearized stability analysis to simplified models of
accretion-phase neutrino fluxes streaming from a supernova. We
compare the results with recent numerical studies and find excellent
agreement. This provides confidence that a linearized stability
analysis can be further applied to more realistic models.
\end{abstract}

\section{Introduction}

Neutrino-neutrino interactions cause the neutrino flux evolution
close to a supernova (SN) core to be nonlinear and numerically very
challenging~\cite{Duan:2010bg}. The flavor instability causing
collective flavor conversions can be suppressed by the ``multi-angle
matter effect'' \cite{EstebanPretel:2008ni}. This point was recently
investigated numerically for an accretion-phase model where the
matter density near the neutrino sphere is large, using a schematic
description of the neutrino fluxes~\cite{Chakraborty:2011gd}. On the
other hand, the flavor stability can also be investigated with a
linearized stability analysis, avoiding an explicit solution of the
equations of motion~\cite{Banerjee:2011fj}. We apply this method to
the models of Ref.~\cite{Chakraborty:2011gd} and find excellent
agreement of the stable regime identified with either method.

\section{Linearized stability analysis}

We describe the neutrino flavor evolution in terms of matrices
${\sf\Phi}_{E,u,r}$ where the diagonal elements are the usual total
number fluxes and the off-diagonal elements encode phase
information~\cite{EstebanPretel:2008ni,Sarikas:2011am}.
We label the angular dependence with $u$, in close
relation with the neutrino emission angle $\vartheta_R$ at the inner
boundary radius $R$,
$u=\sin^2\vartheta_R=(1-\cos^2\vartheta_r)\,r^2/R^2$. For
semi-isotropic emission at a ``neutrino sphere'' with radius $R$,
the flux is uniformly distributed on $0\leq u\leq1$. The equations
of motion are
$
\I\partial_r{\sf\Phi}_{E,u,r}=[\sH_{E,u,r},{\sf\Phi}_{E,u,r}]
$,
with the Hamiltonian~\cite{Banerjee:2011fj}
\beq \nonumber
\sH_{E,u,r} = \left(\frac{\sM^2}{2E}+
\sqrt{2}\,G_{\rm F}\sN_\ell\right)\,\frac{1}{v_{u,r}}
+\frac{\sqrt{2}\,G_{\rm F}}{4\pi r^2}
\int_0^1 \D u'\int_{-\infty}^{+\infty}\D E'
\left(\frac{1}{v_{u,r}v_{u',r}}-1\right)\,{\sf\Phi}_{E',u',r}\,,
\eeq
where $\sM^2$ is the neutrino mass-squared matrix, $\sN_\ell$ the
matrix of net charged-lepton densities which in the flavor basis is
$\sN_\ell={\rm diag}(n_{e}{-}n_{\bar e},n_{\mu}{-}n_{\bar\mu},
n_{\tau}{-}n_{\bar\tau})$ and $v_{u,r}$ is the radial projection of neutrino velocity at the radius $r$. 
Antineutrinos are represented through negative-energy modes ($E<0$)
and negative negative fluxes in the matrices ${\sf\Phi}_{E,u,r}$.
This sign convention simplifies the formalism and obviates any
distinction between neutrinos and antineutrinos.

Henceforth we drop
the explicit subscript $r$ to denote the $r$-dependence of all
quantities.
In the two flavor scenario one can write:
\beq \nonumber
{\sf\Phi}_{E,u}= \frac{{\rm Tr}\,{\sf\Phi}_{E,u}}{2} +\frac{F^e_{E,u,R}-F^x_{E,u,R}}{2}\,{\sf S}_{E,u} \; ,
\eeq
where $F_{E,u}^{e,x}$ are the differential neutrino fluxes at the inner boundary radius $R$
for the $e$ and $x$ flavors.
The flux summed over all flavors, ${\rm Tr}\,{\sf\Phi}_{E,u}$, drops out of the equations of
motion and is conserved in our free-streaming limit.
The ``swapping matrix'' $${\sf S}_{E,u} = \begin{pmatrix} s_{E,u}&S_{E,u}\\S^*_{E,u}&-s_{E,u} \end{pmatrix}\,, $$ encodes
the flavor evolution with
initial conditions $s=1$ and $S=0$.

We expand the Hamiltonian for large
distances from the core and small mixing angle. After dropping its trace we find
\begin{eqnarray} \nonumber
H^{\rm vac}_{E,u} &=& \frac{{\sf M}^2}{2E}\, v_u^{-1} \rightarrow \pm \, \frac{\omega}{2}\,
\begin{pmatrix}
\cos{2\theta}&\sin{2\theta}\\
\sin{2\theta}&-\cos{2\theta}
\end{pmatrix} \, v_u^{-1} \rightarrow \pm \frac{\omega}{2}\,
\begin{pmatrix}1&0\\0&-1\end{pmatrix}\, \left(1+\frac u2 \, \frac{R^2}{r^2}\right)
\,,
\\ \nonumber
H^{\rm m}_{E,u} &=& \sqrt{2}\,G_{\rm F}\,\begin{pmatrix}n_{e}{-}n_{\bar e}&0\\0&0\end{pmatrix}\, v_u^{-1}
\rightarrow \frac{\tilde\lambda}{2} \,
\begin{pmatrix}1&0\\0&-1\end{pmatrix}\, \left(1+\frac u2 \, \frac{R^2}{r^2}\right)\,,\\ \nonumber
H^{\nu \nu}_{E,u} &\rightarrow & \frac{\sqrt{2}\,G_{\rm F}\, R^2}{4\pi r^4}
\int_0^1 \D u'\,
\frac{u+u'}{2}\,\int_{-\infty}^{+\infty}\D E'\, \frac{F^e_{E,u,R}-F^x_{E,u,R}}{2} \,{\sf S}_{E',u'}\,\,.
\end{eqnarray}
where $\tilde\lambda = \sqrt{2}\,G_{\rm F}\,(n_{e}{-}n_{\bar e})$.
We write the neutrino-neutrino part concisely in the form
$H^{\nu \nu}_{E,u} \equiv \mu_r
\int_0^1 \D u'\,\frac{1}{2}(u+u')\,\int_{-\infty}^{+\infty}\D E'\, g_{E,u} \,{\sf S}_{E',u'}$,
where $\mu_r=\mu_R\,R^4/2 r^4$ encodes the strength of the neutrino-neutrino interaction with the parameter
$\mu_R=\sqrt{2}\,G_{\rm F}(F^{\bar\nu_e}_R-F^{\nu_x}_R)/4\pi R^4$.
We further define the dimensionless flavor difference spectrum
$g_{E,u}=(F^e_{E,u,R}-F^x_{E,u,R})/(F^{\bar\nu_e}_R-F^{\nu_x}_R)$
with the normalization in the antineutrino sector
$\int_{-\infty}^0 \D E \int_0^1 \D u \, g_{E,u}=-1$.
The integration over neutrinos (positive energies) gives $\int_0^{\infty} \D E \int_0^1 \D u \, g_{E,u}=(F^{\nu_e}_R-F^{\nu_x}_R)/(F^{\bar\nu_e}_R-F^{\nu_x}_R) \equiv 1+\varepsilon$, with $\varepsilon$ being asymmetry of the spectra.

Next we expand the equations in the small-amplitude limit $|S|\ll1$ which implies, to linear order,
$s=1$. After switching to the variable $\omega=\Delta m^2/2E$ for the energy modes one finds~\cite{Banerjee:2011fj}
\beq \nonumber
{\rm i}\partial_r S_{\omega,u} =
\left[\omega+u(\lambda+\varepsilon\mu)\right]S_{\omega,u}
- \mu \int \D u'\,\D \omega'\,(u+u')\,g_{\omega'u'}\,S_{\omega',u'}\,.
\eeq
Here $\lambda = \tilde \lambda R^2/2r^2$ encodes the imprint of multi-angle matter effect.
Except for the additional two powers of $r^{-1}$ this quantity describes the SN density profile and scales approximately as
$\mu_r\propto r^{-4}$.

Writing solutions of the linear differential equation in the form
$S_{\omega,u}=Q_{\omega,u}\,e^{-{\rm i}\Omega r}$ with complex
frequency $\Omega=\gamma+{\rm i}\kappa$ and eigenvector
$Q_{\omega,u}$ leads to the eigenvalue equation
\cite{Banerjee:2011fj},
\begin{equation}\nonumber
(\omega + u \bar\lambda - \Omega)\, Q_{\omega,u}=
\mu \int du'\,d\omega'\,(u+u')\,g_{\omega'u'}\,Q_{\omega',u'}\,,
\end{equation}
where $\bar\lambda \equiv \lambda + \varepsilon \mu$. The solution
has to be of the form
$Q_{\omega,u}=(A+Bu)/(\omega+u\bar\lambda-\Omega)$. Solutions exist
if $\mu^{-1}=I_1\pm\sqrt{I_0I_2}$, where $I_n=\int
d\omega\,du\,g_{\omega,u}\,u^n/(\omega+u\bar\lambda-\Omega)$. The
system is stable if all $\Omega$ are purely real.
A possible imaginary part, $\kappa$, is the exponential growth rate.

\section{Results}

We aim at comparing the linearized stability analysis with the
numerical solutions of Ref.~\cite{Chakraborty:2011gd} who
numerically solved the neutrino flavor evolution for a $10.8\,
M_\odot$ model at various post bounce times. They confirmed the
multi-angle matter suppression of self-induced flavor conversion,
but also found partial conversions at a large radius for the models
$200\,{\rm ms} \lesssim t_{\rm pb} \lesssim 300\,{\rm ms}$.

\begin{figure}[thb]
\centerline{
\includegraphics[width=0.35 \textwidth, trim = 0 0 0 8mm]{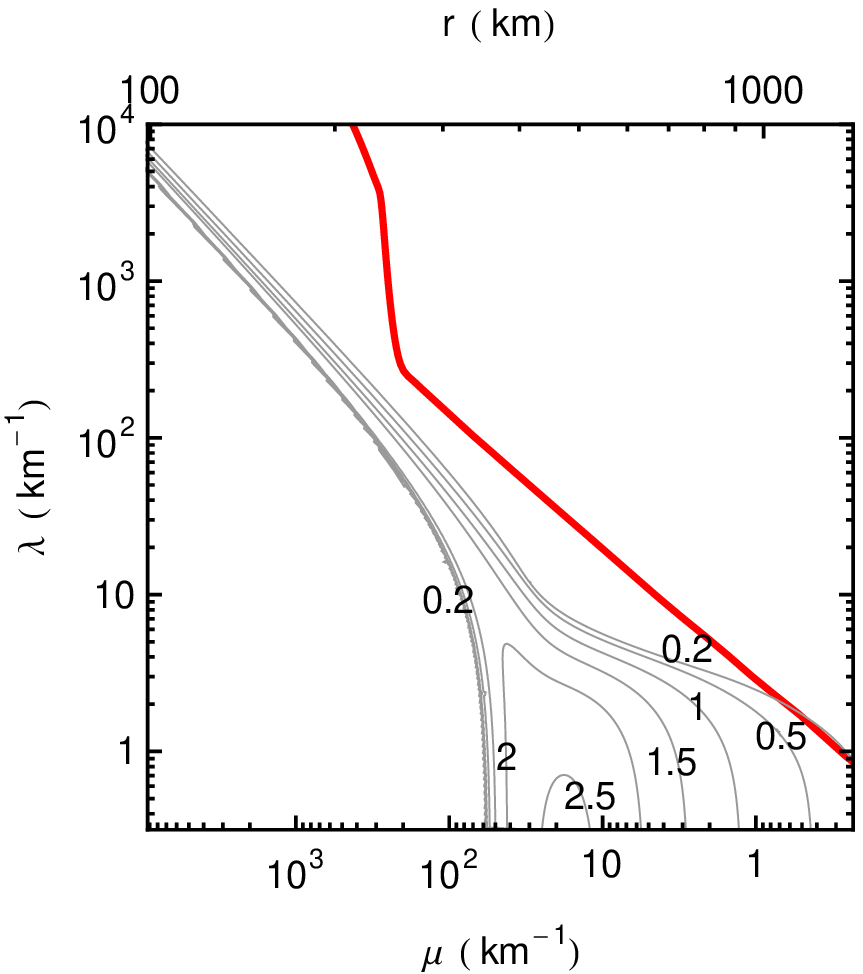}
\includegraphics[width=0.35 \textwidth, trim = 0 0 0 8mm]{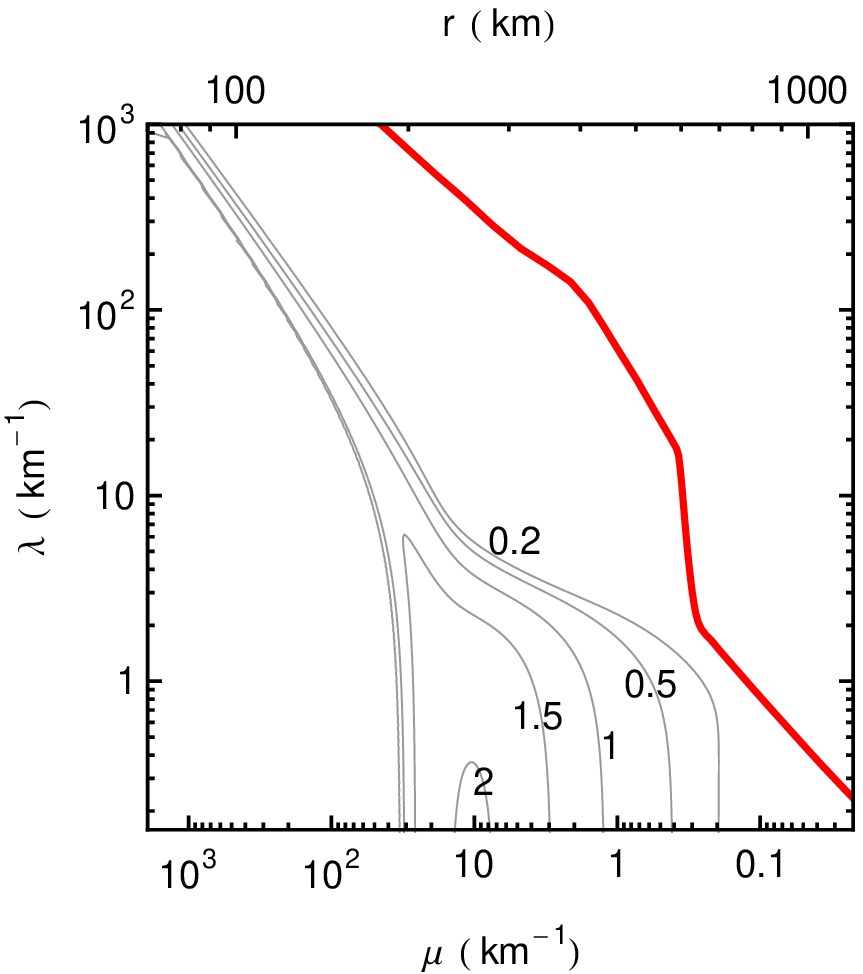}
}
\caption{Contours of $\kappa$ and the trajectory of SN (thick red line) at $t=300$~ms (left) and 400~ms (right) post bounce for a 10.8$\, M_\odot$ model discussed in Ref.~\cite{Chakraborty:2011gd}.}
\label{fig:trajectory}
\end{figure}

We use the same schematic half-isotropic and monochromatic spectra,
leading to the simple form $g_{\omega,u} = -\delta (\omega+\omega_0)
+ (1+\varepsilon) \, \delta (\omega-\omega_0) $. The integrals $I_n$
can now be evaluated analytically. Then it is easy to find a
solution $(\gamma, \kappa)$ for each pair $(\mu, \lambda)$.
Figure~\ref{fig:trajectory} shows the region where $\kappa \neq 0$
for two snapshots together with the $\kappa$ isocontours. We also
show the ``SN trajectory'' in the $(\mu,\lambda)$ plane, i.e.\
essentially the density profile as a function of radius because
$\mu_r\propto r^{-4}$.

Our results agree with the numerical solutions of
Ref.~\cite{Chakraborty:2011gd} for all models. Whenever the
numerical solutions find no flavor conversion, our SN trajectory
indeed stays clear of the unstable regime. Conversely, when it
briefly enters the unstable regime as in the left panel of
Fig.~\ref{fig:trajectory}, we reproduce the onset radius for partial
flavor conversion of Ref.~\cite{Chakraborty:2011gd}. The linear
stability analysis correctly explains the numerical results.

\begin{wrapfigure}{h!}{0.5 \textwidth}
\vspace{-40pt}
\center{
\includegraphics[width=0.4 \textwidth, trim = 0 0 0 0cm]{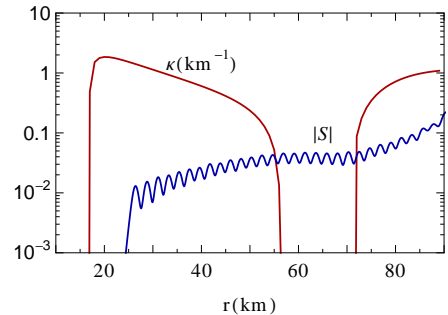}}
\caption{Growth rate $\kappa$ and off-diagonal element $|S|$ for a
toy model (see text).}\vspace{-20pt} \label{fig:peculiar}
\end{wrapfigure}

It is interesting that in principle the SN trajectory can enter the
instability region twice. As a toy model we consider the density
profile $\lambda \sim 0.43 \, \mu $ with half-isotropic emission at
$R=10$~km and $\mu_r = 7 \times 10^4~{\rm km}^{-1}\,{R^4}/{2 r^4}$.
In Fig.~\ref{fig:peculiar} we show $\kappa(r)$ and the evolution of
the off-diagonal element $|S|$. Indeed $|S|$ oscillates and grows in
the unstable regime, only oscillates when $\kappa=0$, and then grows
again during the second instability crossing. It remains to be seen
if there are realistic density profiles where such a multiple
instability situation exists in practice.

\section{Conclusions}

The nonlinear neutrino flavor evolution in the SN environment can be
a challenging numerical task even when it only consists of
post-processing the output of a self-consistent SN simulation, not
to mention solving self-consistently the multi-flavor neutrino
transport. The latter is not necessary if collective oscillations do
not happen in the critical region below the shock wave. The question
if a given SN model with concomitant neutrino fluxes is stable
against self-induced flavor conversion can be answered with a
linearized stability analysis~\cite{Banerjee:2011fj}. Of course, if
the model is unstable, one needs to solve the equations numerically
to find the final outcome. However, since neutrino fluxes during the
accretion phase may well be stable because of the multi-angle matter
effect, a linearized flavor stability analysis is here a useful
tool.

We have applied this method to the models studied in
Ref.~\cite{Chakraborty:2011gd} and compared with the outcome of
their numerical solutions. The results are very encouraging in that
we can perfectly account for the results of the numerical
simulations and can also predict the onset radius for those cases
where partial flavor conversion occurs at a large radius.

Meanwhile we have applied this method to an accretion-phase model
with realistic energy and angle distributions~\cite{Sarikas:2011am}.
We find that a stability diagram in the form of our
Fig.~\ref{fig:trajectory} is an excellent tool to summarize the
flavor stability situation of SN neutrino fluxes.

\section{Acknowledgments}

We thank Sovan Chakraborty and collaborators for providing us their
SN models and Irene Tamborra and Javier Redondo for comments on the
manuscript. We also thank the organizers of the HANSE 2011 workshop
for a very informative meeting and the opportunity to present our
results on very short notice. Partial support by the Deutsche
Forschungsgemeinschaft under grants TR 27 and EXC 153 is
acknowledged.



\begin{footnotesize}

\end{footnotesize}



\begin{thebibliography}{99}


\bibitem{Duan:2010bg}
  H.~Duan, G.~M.~Fuller and Y.-Z.~Qian,
  ``Collective neutrino oscillations,''
  Ann.\ Rev.\ Nucl.\ Part.\ Sci.\  {\bf 60} (2010) 569.

\bibitem{EstebanPretel:2008ni}
  A.~Esteban-Pretel, A.~Mirizzi, S.~Pastor, R.~Tom\`as, G.~G.~Raffelt,
  P.~D.~Serpico and G.~Sigl,
  ``Role of dense matter in collective supernova neutrino transformations,''
  Phys.\ Rev.\ D {\bf 78} (2008) 085012.

\bibitem{Chakraborty:2011gd}
  S.~Chakraborty, T.~Fischer, A.~Mirizzi, N.~Saviano and R.~Tom\`as,
  ``Analysis of matter suppression in collective neutrino oscillations during the supernova accretion phase,''
  Phys.\ Rev.\  D {\bf 84} (2011) 025002.

\bibitem{Banerjee:2011fj}
  A.~Banerjee, A.~Dighe and G.~G.~Raffelt,
  ``Linearized flavor-stability analysis of dense neutrino streams,''
  Phys.\ Rev.\ D {\bf 84} (2011) 053013.

\bibitem{Sarikas:2011am}
  S.~Sarikas, G.~G.~Raffelt, L.~H\"udepohl and H.-T.~Janka,
  ``Flavor stability of a realistic accretion-phase supernova neutrino flux,''
  arXiv:1109.3601.

\end{thebibliography}
\end{document}